# A Novel Self-Packaged DBBPF With Multiple TZs for 5G Sub-6 GHz Applications

Hao Zhang, *Student Member, IEEE;* Kaixue Ma, *Senior Member, IEEE;* Wenwen Zhang, *Student Member, IEEE*

*Abstract*—In this paper, we proposed a self-packaged dual-band bandpass filter (DBBPF) with high isolation and low insertion loss (IL) for 5G sub-6 GHz applications. To get high stopband suppression, multiple and controllable transmission zeros (TZs) are introduced. A pair of quarter-wavelength stepped-impedance resonators (QSIRs) and half-wavelength hairpin resonators (HWHR) are introduced to form cross-coupled five-star topology to generate two passbands and multiple TZs analyzed by mixed EM couplings. A pair of U-shape resonators are designed on G6 to fully excite the resonators and introduce source/load TZs at the same time. The generation of TZs by separate electric and magnetic coupling paths (SEMCP), is discussed. The DBBPF achieves a low IL of 0.85/1.13 dB with fractional bandwidths (FBW) of 11.0%/6.9% at center frequencies of 3.45 GHz and 4.9 GHz for 5G sub-6 GHz application, respectively. The total size is only $0.32\lambda_g \times 0.45\lambda_g$.

*Index Terms*— 5G, dual-band bandpass filter (DBBPF), self-packaged, separate electric and magnetic coupling paths (SEMCP), substrate integrated suspended line (SISL), transmission zeros (TZs).

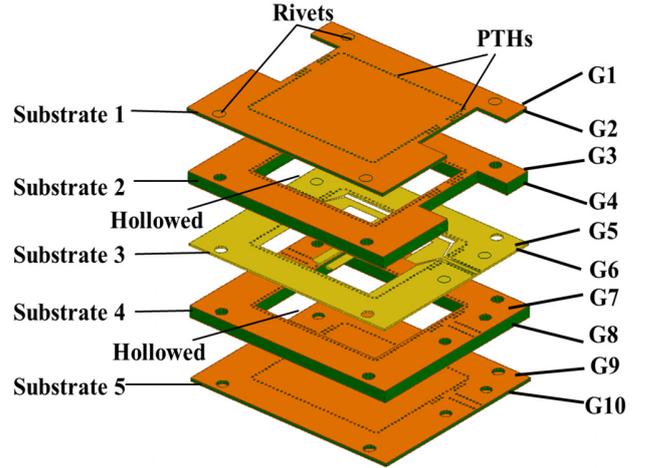

Fig. 1. Proposed SEMCP DBBPF 3-D structure.

## I. INTRODUCTION

WITH the rapid development of wireless communication technology, 5th-Generation (5G) communication technology has gained more and more attention. To develop the 5G communication system, China's Ministry of Industry and Information Technology (CMIIT) released the frequency bands of sub-6 GHz, which cover frequency ranges of 3.3-3.6 GHz and 4.8-5.0 GHz. For better integration, dual-band bandpass filters (DBBPF) for 5G are widely researched which requires low insertion loss and compact size. In previous studies, different kinds of DBBPFs have been proposed. Designs using waveguide resonators [1], substrate integrated waveguide (SIW) [2][3] and microstrip lines, such as ring resonators [4], stepped impedance resonators (SIRs) [5], quad-mode resonators [6], half-wavelength SIR [7] and quarter-wavelength SIRs (QSIRs) [8] are proposed. Especially, excellent out-of-band performance and high skirt selectivity, which can significantly improve by introducing transmission zeros (TZs), are important to the filter design. So, it is important to design a filter with multiple and controllable TZs.

In this letter, a novel DBBPF with multiple TZs designed by using substrate-integrated suspended line (SISL) technology proposed in our patent [9]. The DBBPF for 3.45/4.9 GHz based on cross-coupled five-star topology is composed of a pair of QSIRs and half-wavelength hairpin resonator (HWHR) and excited by a pair of suitable U-shape resonator (also acted as a pair of the feed lines). In particular, besides three TZs in the upper and lower sideband, three more TZs are introduced between the two passbands to improve isolation.

## II. DESIGN OF THE DUAL-BAND BAND PASS FILTER

The structure is shown in Fig. 1, substrates 1 to 5 are covered with ten metal layers on both sides of each substrate, which are termed as G1-G10 respectively. Substrate 3 uses Rogers RT/duroid 5880 ™ with a thickness of 0.254 mm, while Substrate 1,2,4 and 5 use low-cost FR4 epoxy. The thickness of substrates 1 and 5 is 0.6 mm, while the thickness of substrates 2 and 4 is 2 mm. To take advantage of the SISL's double wiring, the circuits are designed on G5 and G6 respectively. The DBBPF is based on SISL technology's advantage of electromagnetic shielding by utilizing metal-plated through holes (PTHs). PTHs surround the hollowed air cavities and aim to reduce electromagnetic loss. If it is not pressed tightly, it will cause electromagnetic wave leakage and affect the frequency response of the circuit. Also, PTHs act as the ground for this DBBPF.

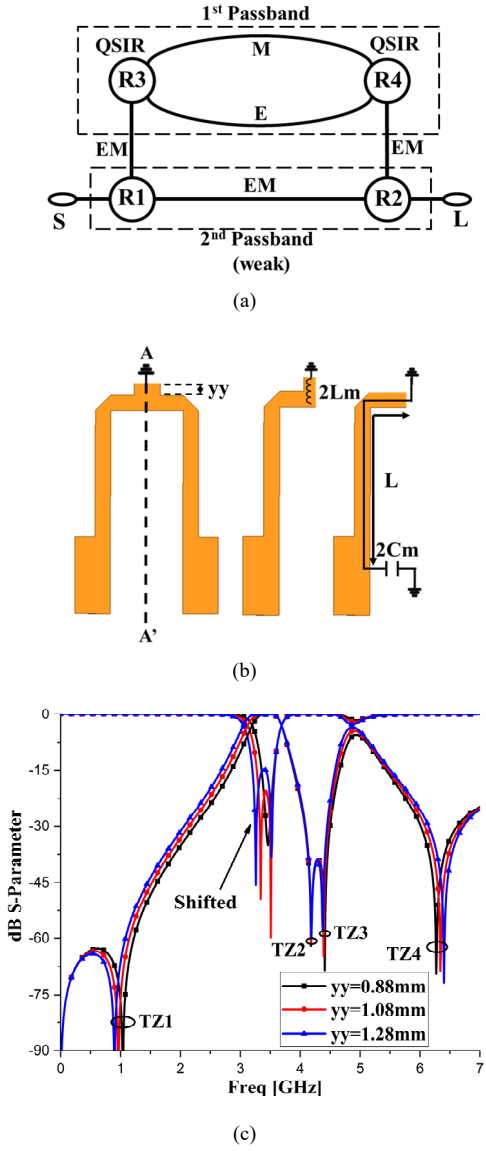

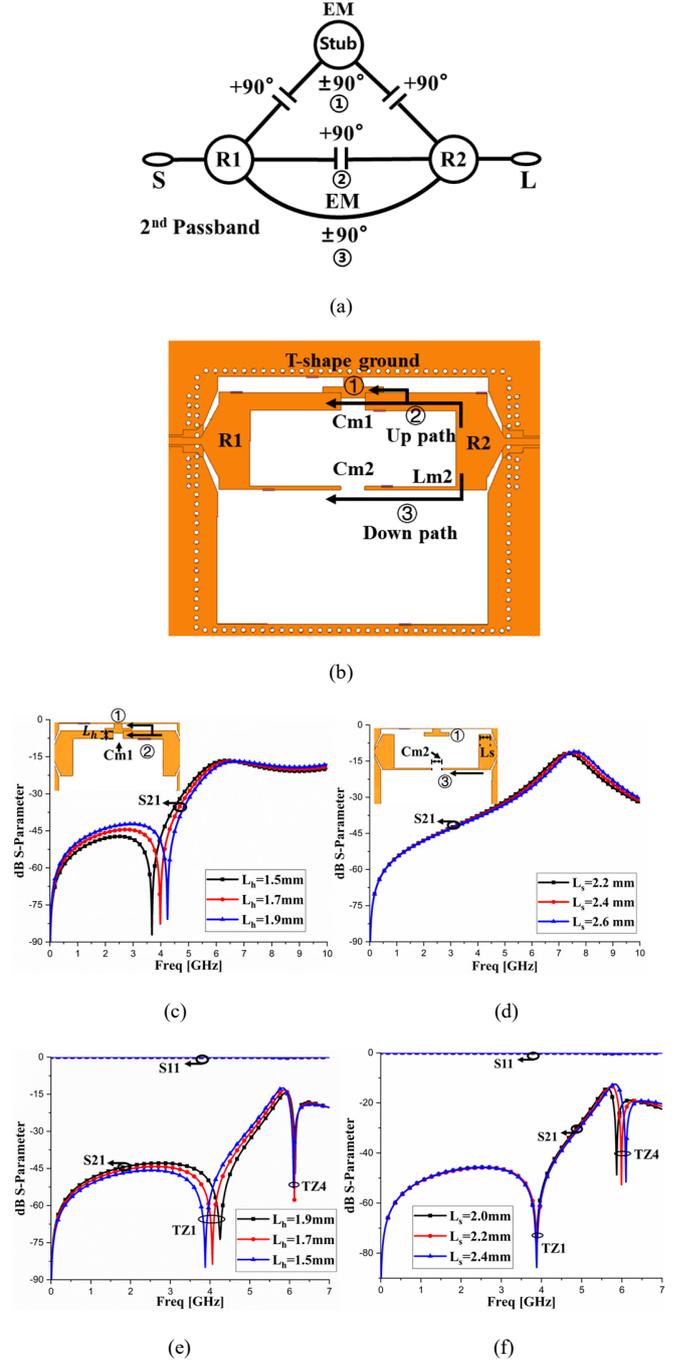

Fig. 2. (a) Primary topology of QSIRs and U-shape half-wavelength resonators. (b) Even and odd mode of QSIRs. (c) Frequency response against yy.

## A. Analysis of the first topology

The proposed DBBPF is initially made of a pair of U-shape half-wavelength resonators called R1/R2 and QSIRs named R3/R4. R1 and R2 are located on the G5 layer while R3 and R4 are located on the G6 layer [10] as the topology shown in Fig. 2(a). This structure forms the prototype of the DBBPF. The resonance of R3 and R4 forms the first passband while the resonance of R1 and R2 forms the second band, as the coupling relationships shown in Fig. 2 (a). The U-shape resonator offers a larger overlapping area between the feed line and the resonator. The advantage of the double-layer wiring in the SISL structure is adopted so that it can be more flexible to design the resonators with different coupling relationships and reduce the circuit size.

This QSIRs structure can be analyzed by even-mode and

Fig. 3. (a) The topology of U-shape resonator and T-shape ground stub. (b) Configurations of U-shape resonator and T-shape ground stub. (c) Frequency response of up-path against $L_s$. (d) Frequency response down-path against $L_s$. (e) Frequency response against $L_h$. (f) Frequency response against $L_s$.

TABLE I
THE PHASE SHIFTS OF DIFFERENT PATHS UNDER DIFFERENT FREQUENCIES

| Path | Lower than $f_0$ | Higher than $f_0$ |
|---|---|---|
| ① | +90°+90°+90°=270°=-90° | +90°-90°+90°=+90° |
| ② | +90° | +90° |
| ③ | +90° | -90° |

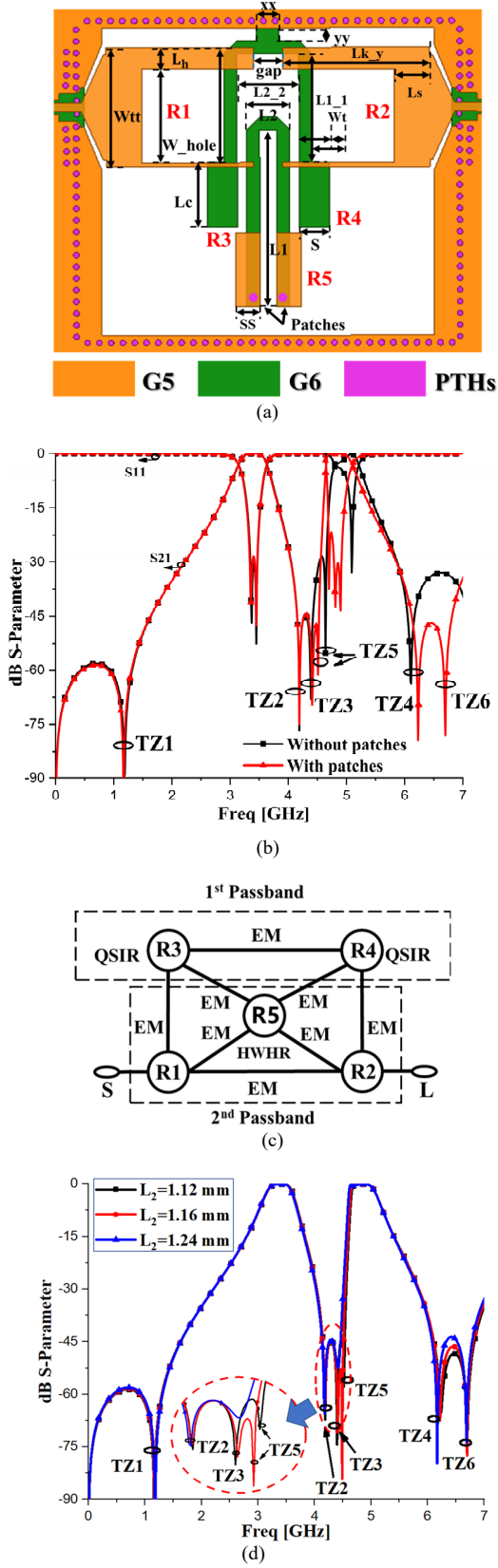

Fig. 4 (a) Proposed filter structure. (b) The frequency responses with/without the patches. (c) The cross coupled five-star topology. (d) The frequency response against $L_2$.

the equivalent capacitance of QSIRs, which is called the E coupling path. $L_m$ is the equivalent inductance of the shorted transmission line, which forms the M coupling path. M and E coupling are controlled by $L_m$ and $C_m$. According to the mechanism of resonance [12], when the input admittances are zero for both the even and odd modes, the structure is resonant. The frequencies of even and odd modes are given by [11].

$$f_e = \omega_e/2\pi = 1/4(2Y_c L_m + \sqrt{\varepsilon_{re}}L/c) \quad (1)$$

$$f_o = \omega_o/2\pi = 1/4(2Z_c C_m + \sqrt{\varepsilon_{re}}L/c) \quad (2)$$

$L$ is the length of the transmission line, $Z_c$ is the characteristic impedance of the resonator and $c$ is the light speed in the free space. The frequency $f_0$ is determined by $f_0 = (f_e + f_o)/2$. The coupling coefficient $C$ is concluded as:

$$C = (\omega_O^2 - \omega_e^2)/(\omega_o^2 + \omega_e^2) = FY_C L_m - FC_m Z_c = M - E \quad (3)$$

$$F = (4(\sqrt{\varepsilon_{re}}L/c + Y_c L_m + C_m Z_c)/(2Y_c L_m + \sqrt{\varepsilon_{re}}L/c)^2 \quad (4)$$

In this structure, $L_m$ is controlled by stub yy of QSIRs. As Fig. 2 (c) shows, by increasing the length of the short stub yy (T-shape ground stub connects with the ground), $L_m$ decreases, which leads to the decrease of $f_e$ and the increase of FBW, while $f_o$ keep still. When $C_m > 0$, it can generate an additional TZ (TZ2). TZ3 is produced by the harmonic effect [10][11][13].

TZ1 and TZ4 are also introduced by SEMCPs. As is shown in Fig. 3 (a) and (b), when only R1, R2, and T-shape ground stub are retained, three signal paths exist ① S-R1-Stub-R2-L; ② S-R1-R2-L (upper half); ③ S-R1-R2-L (lower half). When only path ② remains, TZ1 is generated, shown in Fig. 3 (c).

When only path ① remains, no TZ is generated, shown in Fig. 3 (d). Therefore, it can be concluded that the effect of path ① with a T-shape ground stub is influenced by path ② and ③. When the resonance frequency is lower than the center frequency of the passband, TZ1 is generated by the canceling of M coupling in ① and the E coupling in ②. When the resonance frequency is higher than the center frequency of the passband, TZ4 is generated by the canceling of E coupling in ① and the M coupling in ③. It can be seen that the coupling types in ① and ③ are mixed EM coupling and frequency-dependent [14]. Because the lower part in ③ is of high impedance, which appears as inductance $L_m 2$, while the gap $C_m 2$ in this part functions as capacitance. $C_m 2$ and $L_m 2$ leads to the mixed EM coupling.

When the signal frequency is lower than the resonant frequency, capacitance is dominant and the coupling type in path ③ is E coupling, which leads to the +90° phase shift. When the signal frequency is higher than the resonant frequency, inductance is dominant and the coupling type in path ③ is M coupling, which leads to a -90° phase shift [14][15].

Similarly, the T-shape ground stub in ① is M dominant and works as inductance, while the $C_m 1$ nearby appears as odd-mode with reference line A-A', shown in Fig. 2 (b). $C_m$ is capacitance. The phase shift in this structure is displayed in Table I. If increasing $L_h$, $C_m1$ can be enhanced, and TZ1 moves to the working frequency. Increasing $L_s$ can change the M coupling, and TZ4 moves to a higher frequency. Thus, as shown in Fig. 3 (e) - (f), we can control TZ1 or TZ4 by changing $L_s$ or $L_h$, separately. Since we separated the whole structure into several parts to analyze, the physical length of the resonator is reduced, its resonant frequency moves to high frequency. So, the corresponding positions of TZs are not matched as the same as Fig. 2 (c).

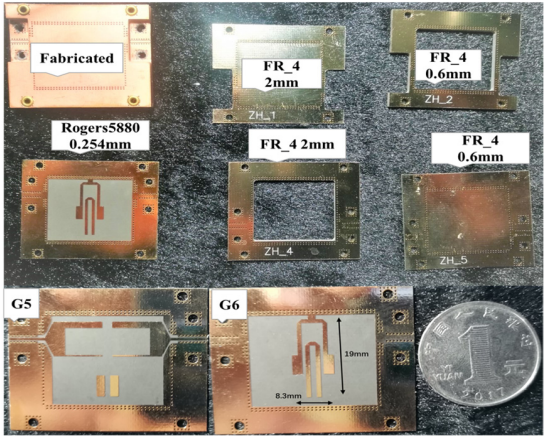

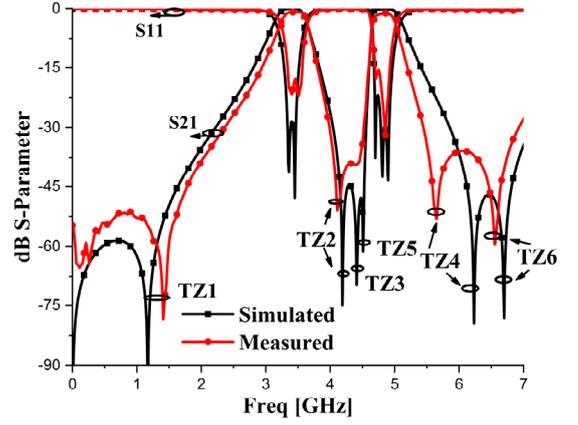

Fig. 5. Photograph of Fabricated DBBPF with HWHR. (a) Five layers and compacted filter. (b) G5 and G6.

Fig. 6. Simulated and measured results of the DBBPF.

TABLE II
COMPARISON WITH REPORTED WORKS

| Ref. No. | Types | $f_0$ (GHZ) | FBW (%) | IL (dB) | TZs | Self-packing | Size ($\lambda_g * \lambda_g$) |
|---|---|---|---|---|---|---|---|
| [4] | Microstrip | 2.4/5.8 | 12.0/12.0 | 1.40/3.20 | 2 | No | 0.26*0.13 |
| [5] | Microstrip | 1.57/2.38 | 9.1/6.5 | 1.21/1.95 | 3 | No | 0.62*0.39 |
| [17] | Microstrip | 1.78/4.0 | 7.0/5.5 | 0.42/0.37 | 4 | No | 0.23*0.33 |
| [18] | Microstrip | 2.4/3.57 | 7.5/5.0 | 0.87/1.90 | 1 | No | 0.50*0.20 |
| [19]-1 | SISL | 2.55/3.97 | 12.5/5.4 | 1.59/2.91 | 3 | Yes | 0.15*0.14 |
| [19]-2 | SISL | 2.55/3.91 | 13/5.4 | 2.0/3.4 | 3 | Yes | 0.15*0.15 |
| [20] | SISL | 3.45/4.9 | 7.82/4.08 | 1.15/1.42 | 5 | Yes | 0.21*0.31 |
| **This work** | **SISL** | **3.45/4.9** | **11.0/6.9** | **0.85/1.13** | **6** | **Yes** | **0.32*0.45** |

### B. Analysis of the second topology with HWHR

R1 and R2, with half-wavelength electric length at 4.9 GHz, generate the second passband. At 4.9 GHz, the coupling between R3 and R4 is relatively weak, since they do not resonant but function as loading elements of R1 and R2 in this circumstance. To strengthen the coupling of the second passband, HWHR, with a half-wavelength electric length at 4.9 GHz, is designed on the G6 layer. From the result, HWHR greatly improves the performance of the second passband without any influence on the first passband.

Especially, as is shown in Fig. 4(a), the electric coupling part of HWHR is folded into the G5 layer connected with PTHs. This could lead to miniatured circuit size and strong E coupling by this pair of patches on the G5 layer, as is shown in Fig. 4(b). By introducing HWHR, the primitive weak E coupling path of QSIRs is changed into the mixed EM coupling path. The cross-coupled five-star topology of this filter is shown in Fig. 4(c). The M path is realized by L2 and the E path is realized by the folded patches on G6. Therefore, an additional TZ called TZ5 is generated at the left of the second passband [11] due to SEMCP. Now, five resonators are working simultaneously where R1, R2, and R5 generate the second passband, while R3 and R4 work for the first passband.

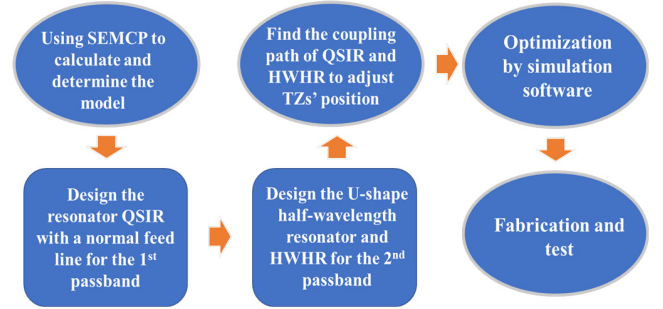

Fig. 7. The design process of the DBBPF in this paper.

As is shown in Fig. 4(d), alter the M coupling path by decreasing of length $L_2$ makes TZ5 move to the higher frequency [11] and finally TZ3 and TZ5 will be resolved [16]. TZ6 is generated by the third passband at around 10 GHz. There are three transmission poles at the second passband introduced by the HWHR, R1, and R2. This filter performance is satisfying for 5G sub 6 GHz.

To design this DBBPF, we followed the design process, as is shown in Fig. 7. The proper parameters for this DBBPF are: $L_1$ =13 mm, $L_2$ =1.14 mm, L1_1 = 8.3 mm, L2_2 = 4.2 mm, Wt = 0.9 mm, S = 1.7 mm, SS = 1.7 mm, $L_s$= 2.2mm, $L_h$= 1.5mm, Lk_y = 8mm, Wtt = 8.2 mm, Lc = 4.5 mm, gap = 2 mm, W_Hole = 6.4 mm, xx =1.52 mm, yy =0.88 mm.

## III. RESULTS

The DBBPF based on SISL has been fabricated and measured. Fig. 5 shows the five substrate layers of SISL. They are bonded together orderly with rivets. The size of the main circuit is $0.21\ \lambda_g \times 0.25\ \lambda_g$. The total size is only $0.32\ \lambda_g \times 0.45\ \lambda_g$ Fig. 6 shows the simulated and measured results of this DBBPF. The fractional bandwidths (FBW) of the two passbands are 11.0%/6.9% with insertion losses of 0.85/1.13 dB. Due to insufficient machining accuracy, the measured TZ3 and TZ5 are resolved as is analyzed in Section *B*. The performance is summarized in Table II. This proposed filter achieves the advantages of low insertion loss, multiple TZs, and self-packaging for the passbands whose center frequencies are 3.45/4.9 GHz.

## IV. CONCLUSION

In this letter, the DBBPF based on SISL is designed and measured. HWHR is specially designed to fully excite the second passband at 4.9 GHz. The DBBPF based on cross-coupled five-star topology has advantages of excellent electromagnetic shielding, easy for wiring, and self-packaging with six TZs. Using the advantages of double-layer wiring in SISL structure, it can not only stimulate the circuit of another layer but also produce another passband. In conclusion, this structure can be potentially applied for the dual-band components and base station for the 5G sub-6 GHz communication system.